\documentclass[acmsmall]{acmart}
\AtBeginDocument{%
  \providecommand\BibTeX{{%
    \normalfont B\kern-0.5em{\scshape i\kern-0.25em b}\kern-0.8em\TeX}}}

\setcopyright{acmcopyright}
\copyrightyear{2025}
\acmYear{2025}
\acmDOI{XXXXXXX.XXXXXXX}

\usepackage{hyperref}

\acmConference[CSCW 2025]{CSCW}{}{}
\acmBooktitle{} 
\acmPrice{}
\acmISBN{}

\newcommand{\yun}[1]{\out{{\small\textcolor{teal}{\bf [*** Yun: #1]}}}}

\newcommand{\si}[1]{\out{{\small\textcolor{red}{\bf [*** Si: #1]}}}}

\begin{document}

\title[Deaf-Inclusive Video-based Learning]{Inclusive Emotion Technologies: Addressing the Needs of d/Deaf and Hard of Hearing Learners in Video-Based Learning}

\author{Si Chen}
\email{sic3@illiois.edu}
\orcid{0000-0002-0640-6883}
\affiliation{%
  \institution{School of Information Sciences, University of Illinois Urbana-Champaign}
  \city{Champaign}
  \state{Illinois}
  \country{USA}
  \postcode{61802}
}

\author{Jason Situ}
\email{junsitu2@illinois.edu}
\affiliation{
  \institution{Computer Science, University of Illinois Urbana-Champaign}
  \city{Urbana}
  \state{Illinois}
  \country{USA}
}

\author{Haocong Cheng}
\email{haocong2@illinois.edu}
\affiliation{
  \institution{School of Information Sciences, University of Illinois Urbana-Champaign}
  \city{Champaign}
  \state{Illinois}
  \country{USA}
}

\author{Suzy Su}
\email{xiaoyus4@illinois.edu}
\affiliation{
  \institution{School of Information Sciences, University of Illinois Urbana-Champaign}
  \city{Champaign}
  \state{Illinois}
  \country{USA}
}

\author{Desirée Kirst}
\email{deskirst@gmail.com}
\affiliation{
  \institution{Gallaudet University}
  \city{Washington}
  \state{District of Columbia}
  \country{USA}
}

\author{Lu Ming}
\affiliation{
  \institution{Gallaudet University}
  \city{Washington}
  \state{District of Columbia}
  \country{USA}
}

\author{Qi Wang}
\email{qi.wang@gallaudet.edu}
\affiliation{
  \institution{Gallaudet University}
  \city{Washington}
  \state{District of Columbia}
  \country{USA}
}

\author{Lawrence Angrave}
\email{angrave@illinois.edu}
\affiliation{
  \institution{Computer Science, University of Illinois Urbana-Champaign}
  \city{Urbana}
  \state{Illinois}
  \country{USA}
}

\author{Yun Huang}
\email{yunhuang@illinois.edu}
\affiliation{
  \institution{School of Information Sciences, University of Illinois Urbana-Champaign}
  \city{Champaign}
  \state{Illinois}
  \country{USA}
}



\begin{abstract}

Accessibility efforts for d/Deaf and hard of hearing (DHH) learners in video-based learning have mainly focused on captions and interpreters, with limited attention to learners' emotional awareness--an important yet challenging skill for effective learning. Current emotion technologies are designed to support learners' emotional awareness and social needs; however, little is known about whether and how DHH learners could benefits from these technologies. Our study explores how DHH learners perceive and use emotion data from two collection approaches, self-reported and automatic emotion recognition (AER), in video-based learning. By comparing the use of these technologies between DHH (N=20) and hearing learners (N=20), we identified key differences in their usage and perceptions: 1) DHH learners enhanced their emotional awareness by rewatching the video to self-report their emotion and called for alternative methods for self-reporting emotion, such as using sign language or expressive emoji designs; and 2) while the AER technology could be useful for detecting emotional patterns in learning experiences, DHH learners expressed more concerns about the accuracy and intrusiveness of the AER data. Our findings provide novel design implications for improving the inclusiveness of emotion technologies to support DHH learners, such as leveraging DHH peer learners' emotions to elicit reflections.

\end{abstract}



\maketitle

\section{Introduction} 

Video-based learning has become increasingly popular, allowing people to learn without being restricted by physical locations and supporting learner-led learning. Video-based learning is an online learning method that relies on live or prerecorded video. For video-based learning to be effective, informed by self-regulated learning (SRL) theory, emotional skills such as being aware of and reflecting on one's emotions and learning process are important \cite{delen2014effects}. SRL involves thoughts, feelings, and behaviors that learners plan and adjusts to achieve learning goals \cite{matcha2019systematic, zimmerman2009self}. When making video-based learning inclusive for d/Deaf and Hard of Hearing (DHH) learners, the focus has often been on improving access to learning materials, typically involving adding well-designed text captions \cite{bhavya2022exploring, kafle2016effect, berke2020deaf} or incorporating sign language captioning \cite{mehta2020automated}. However, there is limited work on how emotional skills can enhance (in)accessibility for learners, and how technology can support these skills.

Encouraging learners with disabilities to set their own goals and assess their own progress has been challenging \cite{lowy2023building,wiener2004social}. Focusing on DHH learners, 90\% of whom are born to hearing parents who have limited experience with deafness or sign language, which may significantly impact their communication, language development, and emotional skills.  Research also shows that DHH learners may struggle to develop emotional skills at the same rate as their hearing peers \footnote{\url{https://www.isbe.net/Documents/Social-Emotional-Learning-Deaf-Hard-Hearing-Students.pdf}}. Based on the United States Census \cite{2022}, an estimated 11.5 million people in the US have various degrees of hearing loss, which makes up 3.5\% of the population. Without understanding and technology support for emotional skills, learners with disabilities are at a higher risk in an education landscape that increasingly emphasizes learner-led approaches. For example, recent studies have found that awareness and understanding of one's thought processes are crucial for using generative AI \cite{tankelevitch2024metacognitive}.

CSCW and HCI research has explored technology support for increasing learners' emotional awareness and reflection in online learning, but few studies have considered how learners with disabilities might or might not benefit similarly. There are two common types of emotion data used in the technology: Automatic Emotion Recognition (AER) and self-reported emotion. AER, a type of emotion technologies \cite{roemmich2021data, kaur2021emotion, mcstay2018emotional} that recognizes, interprets, and categorizes human emotion through multimodal cues, has been visualized in dashboards to help learners become more aware of their emotional experiences (e.g., \cite{chen2022mirror, munoz2020emotion, bahreini2016towards}). In addition to observational AER, self-reported emotion is also used to increase emotion awareness, providing an introspective perspective (e.g., \cite{lbw, chen2022mirror, kiskola2021applying, yan2022emoglass}). To further increase awareness of learners' own emotions, peer learners' emotion was also found to be helpful (e.g.,\cite{mirrorus, bazarova2015social, semsioglu2018emotionscape}).

In studying education for learners with disabilities, there are two common groundings: \textit{Inclusive Education} and \textit{Special Education}. \textit{Inclusive Education} emphasizes transforming mainstream education to educate all learners, while \textit{Special Education} focuses on individual assessment and specialized instruction \cite{salend2010creating}. These two theories are often viewed as diametrically opposed in their approaches, each with its own supporters. Our work builds upon Hornby's theory of \textit{Inclusive Special Education}, which integrates both perspectives \cite{hornby2014inclusive}. This approach intends to ensure effective education for all learners with educational needs, facilitating the highest level of inclusion in society post-school. 

Therefore, we explore whether and how mainstream education interventions can be adapted for DHH learners to include as many learners as possible, using these interventions to promote the emotional skills necessary for SRL post-school. More specifically, we study whether DHH learners can benefit from a video-based learning prototype initially designed for mainstream education to increase emotional awareness and reflection. The prototype collects and supports reflections on two types of emotional data: self-reported emotions and AER. AER technology was used during video-watching to recognize learners' emotion through facial expression, whereas self-reported emotion was collected post-watching. Learners further reflected on two types of their own and peer learners' emotions visualized on a learning analytic dashboard (Figure \ref{fig:flow}). We conducted Zoom-based user studies with DHH (N=20) and hearing (N=20) college students, comparing the use of self-reported emotion, AER, surveys, and interviews to address our research questions: 

\begin{itemize}

\item \textbf{RQ1}: How do DHH learners perceive and use self-reported emotions differently from hearing learners, and why?

\item \textbf{RQ2}: How do DHH learners perceive and use AER-generated emotions differently from hearing learners, and why?

\end{itemize}

Our paper makes the following contributions: (1) we found DHH and hearing learners did not reflect similarly on AER and self-report emotion; (2) we gained an understanding of how current approaches for emotion data, self-reported and AER, are less inclusive for DHH learners and gathered suggestions for improving their designs; (3) we proposed design insights for improving emotion awareness of DHH students, such as leveraging peer learners' emotions to elicit reflections.

\section{Related Works}






\subsection{Emotion Awareness and Social Needs for DHH Learners} 

Prior works highlight the importance of inclusive learning for underserved learners. The Universal Design for Learning (UDL) framework \cite{rose2000universal}, a well-established model, offers actionable recommendations for inclusive education \cite{brandt2022universal}. Derived from Universal Design, UDL aims to create experiences that accommodate individuals with diverse abilities \cite{rose2006universal}. It provides educators with three primary principles: multiple means of representation, multiple means of action and expression, and multiple means of engagement \cite{rose2006universal}. 

Video-based learning is flexible and engaging, and creating inclusive content for DHH learners is important. Captions are commonly used, but their quality is often low due to errors in auto-generation \cite{bhavya2022exploring, kafle2016effect, berke2020deaf}. Tools like ClassTranscribe \cite{angrave2020benefits} improve caption accuracy with crowdsourcing, but they may still be insufficient due to the diverse backgrounds of DHH learners \cite{kushalnagar2010multiple, kushalnagar2014accessibility}. Sign language is an effective alternative, but few resources include it due to high costs and a shortage of proficient interpreters \cite{lang2002higher, stinson2019improving}.
A recent study developed a sign language commenting mechanism in video-based learning to create inclusive and enriched social learning experiences for DHH learners \cite{chen2024towards}.
Systems automated 3D sign language captions show promise, but more improvements are needed for full engagement and inclusivity in video-based learning \cite{mehta2020automated}.

However, increasing content inclusion is insufficient in learning inclusion, among which emotion awareness is also important. Supporting emotional experience and providing multiple means of student engagement are important aspects of UDL. UDL suggested providing options for SRL, an iterative and sequential process that involves multiple phases of learning (e.g., performance, self-reflection, forethought), including monitoring of cognitive, meta-cognitive, behavioral, motivational, and emotional/affective aspects of learning \cite{matcha2019systematic, zimmerman2009self}.  Learners' emotions are also vital for motivation, engagement \cite{pekrun2002academic, linnenbrink2004role}, and academic performance \cite{pekrun2014emotions}.
There are other works on studying SRL from a social perspective about emotions.

Emotional awareness and socialization are important in improving the overall learning experience \cite{feidakis2011endowing, lavoue2020towards} but challenging for learners with disabilities, both offline and online.
Prior studies indicated that assessing students' emotion regulation, emotional skills, and social-emotional competence can provide valuable insights to guide interventions and support strategies, particularly for students in special education\cite{whitcomb2013behavioral, gross2014emotion, jacobs2014emotion}.
Meanwhile, encouraging learners with disabilities to set their own goals and assess their own progress has been challenging \cite{lowy2023building, wiener2004social}.
Peer learners' emotions considered an indirect indicators of well-being for learners with learning disabilities because their moods are often affected by their peers \cite{sharabi2014predictors}. 
Learners with disabilities (e.g. DHH individuals) may have difficulty developing emotional skills with their peers since they have limited access to information learned incidentally and experienced in language deprivation \cite{socialdhh2021}.
Moreover, prior studies \cite{elias2004connection, espelage2016social} found that successful inclusion of learners with disabilities relies largely on the positive peer learners' social-emotional learning.

\subsection{Emotion Awareness Intervention From Mainstream Education}

Researchers often discuss two theoretical approaches for learners with disabilities: \textit{Inclusive Education} and \textit{Special Education}. \textit{Special Education} focuses on individual assessment, planning, and specialized instruction. In contrast, \textit{Inclusive Education} emphasizes educating learners alongside their age peers, and transforming mainstream classrooms. These approaches stem from different philosophies and offer alternative views on educating children with special needs\cite{salend2010creating}. They are increasingly seen as opposing methods, each with its supporters, and it is hard to say which is better. Special education teachers face challenges in choosing suitable formative and summative assessments for students with disabilities, especially online, due to limited theoretical guidance and frequent burnout\cite{cortes2022special}. Mainstream teachers often feel unprepared to teach students with disabilities. Our study builds on the theory of \textit{Inclusive Special Education} \cite{hornby2014inclusive}, combining values from both approaches to facilitate high levels of inclusion post-school. We adapt mainstream education interventions and use so to support learner-led individualized learning. We also believe that future HCI and CSCW research can benefit from situating their work within these approaches, even though it is not commonly discussed yet.

From mainstream education, there are two common types of learners' emotions: self-reported emotion and AER. Self-reported emotion provide direct insight into learners' first-person emotional experiences, while observational methods, such as AER, offer a third-person perspective on emotional expressions. Both AER and self-reported emotion can be leveraged to help learners better understand their emotional experiences. 

\textit{Self-Reported Emotion}. From a learners' first-person perspective, self-reported emotion offers a relatively inexpensive approach \cite{barrett2004feelings} for learners to be more self-aware of their emotional experiences while engaged in a learning process \cite{chen2022mirror, mirrorus, lbw}.
For self-reported emotion, prior works explored using emojis to express emotions in online learning \cite{rong2023understanding, ma2022glancee}.
In the context of video-based learning, Chen et al. \cite{lbw,chen2022mirror} explored how learners reported emotions through emojis and text.
Self-reported emotion also provides a research opportunity to evaluate and improve AER \cite{kaur2022didn}.
Prior works \cite{hilliard2019exploring, han2020students, tzafilkou2021negative} explored how self-reported emotions related to academic performance performance, highlighting its potential as a predictor of student success.
However, self-reported emotions may not always matched observed behaviors, thus causing discrepancies between self-reported and actual emotions \cite{pekruna2020commentary, parry2021systematic}. 

\textit{AER.} AER recognizes, interprets, and classifies human emotions through multimodal and multimedia settings, such as sensing facial expressions, gestures, voices, and bodily behaviors \cite{roemmich2021data, kaur2021emotion, mcstay2018emotional}.
Incorporating AER through facial expressions can enhance system performance and personalization, leading to more precise and customized emotion recognition \cite{dubey2016automatic}. 
For example, when AER has been used in online learning settings with a camera, the results may be affected by lightning or rotation of the facial images \cite{dubey2016automatic}.
Online learning systems (e.g. MetaTutor \cite{azevedo2012metatutor}, \cite{gupta2023facial}), leverage AER technology to enhance the learning experience through fostering self-observation and self-reflection. 
Chen et al. \cite{chen2022mirror} found that by reflecting on visualized AER, learners were able to better reflect on their learning experiences.
Ez-zaouia et al. \cite{ez2017emoda} found that AER dashboard helps tutors to effectively monitor learners' emotions and understand their development during synchronous learning activities. Chen et al. \cite{mirrorus} suggested that seeing peer learners' emotional data can make them emotionally aware of others' learning experiences, and thus fostering their own reflection.

\subsection{Ethical AER for Learner with Disabilities} \label{sec:related_aer}

Prior studies also raised ethical concerns with integrating emotion AI into online learning systems. For example, learners expressed concerns on whether the emotion AI using AER technology could accurately capture their emotion during the learning process \cite{chen2022mirror, reiss2021use, holmes2021ethics, mirrorus}, as they may have different interpretations of their emotion \cite{barrett2019emotional, roemmich2021data, grill2022attitudes}. Learners also raised privacy and transparency concerns \cite{zhao2017semi, kim2020messaging, li2021visual}, as they were unsure how the AI used their facial data and whether the system might share their emotional data with others \cite{murali2021affectivespotlight, samrose2021meetingcoach, chen2022mirror}. Furthermore, emotion AI brought the potential risk of surveillance of both learners and instructors \cite{brown2020seeing,reiss2021use,li2021visual}. However, the current effort in building ethical emotion AI-based online learning systems rarely includes the perspective of underserved learners and understanding how they perceive the ethical concerns of emotion AI, a subfield of AI. 

Among the limited works with emotion AI for underserved populations, Zolyomi and Snyder \cite{zolyomi2021social} studied the emotional experience with emotion AI for neurodivergent learners and made design suggestions for more inclusive emotion AI.
Boyd and Andalibi examined the proposed use of AER technology in the workplace, revealing potential ethical considerations and social implications associated with its implementation \cite{boyd2023automated}.
Shi et al. \cite{shi2022toward} investigated the feasibility of modeling AER to capture the emotions of learners with autism spectrum disorders, thereby paving the way for personalized social systems tailored to individuals' socio-emotional needs.
Our work builds on these works in exploring the design of inclusive and ethical emotion AI using AER technology for underserved learners.

\section{Method}

Our study goals were to understand whether the two approaches of supporting learners' emotional awareness, AER and self-reported emotion, and reflecting on the two could benefit the same for two groups of college students with diverse hearing abilities, one with 20 DHH college students (DHH group) and the other with 20 hearing college students (hearing group), using the same video-based learning prototype. In the rest of the paper, we refer to the DHH participants as D1 to D20, and the hearing participants as H1 to H20. 
Facial expression of participants were captured during video-watching using AER technologies.
After participants watched the video, participants' self-reported emotion were recorded, and then participants observed their own and peers' emotion presented in a dashboard. The study processes were the same among DHH and hearing participants, but we provided multiple communication options to accommodate DHH participants' preferences. Each hearing participant's study session lasted for around one hour and thirty minutes, while DHH participant's study sessions ranged from two hours to two hours and thirty minutes. Participants received compensation of \$15 per hour for their time and contribution involved in taking part in the study. 
This study was approved by the Institutional Review Board.

\textbf{Positionality of Research Team}
Our research team comprised members with diverse hearing status, including hearing and Deaf, HoH individuals: the leading author is hearing and an active American Sign Language (ASL) learner who designed, attended and led analysis in both DHH and hearing's user studies. The DHH's user studies were conducted 
between Fall 2022 and Fall 2023, and hearing participants' user studies were conducted in Fall 2023. The leading author and two other hearing team members conducted the hearing participants' studies. For the DHH participants' user studies, two undergraduate DHH research assistants (RAs) and a graduate hearing RA who is also an ASL interpreter took turns leading user studies. 
Each study session with DHH participants was mentored by a hearing professor with 30 years of experience in teaching DHH students at college level. Two hearing team members did not conduct any of the studies and only joined the team members mentioned above for analysis of codes and themes on a weekly basis. During our weekly meetings, DHH and hearing members communicated with the assistance of ASL interpreters or with auto-captions on Zoom depending on DHH members' preferences. 


\subsection{Participant Recruitment}
For the DHH group, participants were recruited in a university for DHH students in the United States by distributing study information via email lists of a undergraduate-level course, as well as posting recruitment flyers on the college campuses.
In total, 20 DHH college students were recruited in our study.
Out of a total of 20 participants, 12 self-identified them as male, seven as female, and one as non-binary.
Five participants (D2, D12, D13, D14, D17) were identified as hard-of-hearing (HoH), while the other 15 participants were identified as Deaf. As for their first language, ten chose English, four chose other spoken language, eight chose ASL, and two chose other sign language. Note that four participants chose more than one language as their first language.
For the most comfortable classroom instruction language, eight participants chose ASL, three chose English, and the other nine chose both ASL and English.

For the hearing group, we recruited participants through posting flyers on college campuses and other public locations.
We shared recruitment messages in online social media sites like Reddit or via emails to all students of two undergraduate-level classes.
In the recruitment message, we included a survey form for participants to sign up and explicitly stated to seek undergraduate students with minimal knowledge and experience in Augmented Reality, as they would be given a video to learn about it. A total of 20 current or recently graduated undergraduate students in the US were recruited.
Out of a total of 20 participants, 13 self-identified as male, and seven self-identified as female. All hearing participants used spoken language as their first language.


\begin{figure*}[ht]
    \centering  
    \includegraphics[width=1\textwidth]{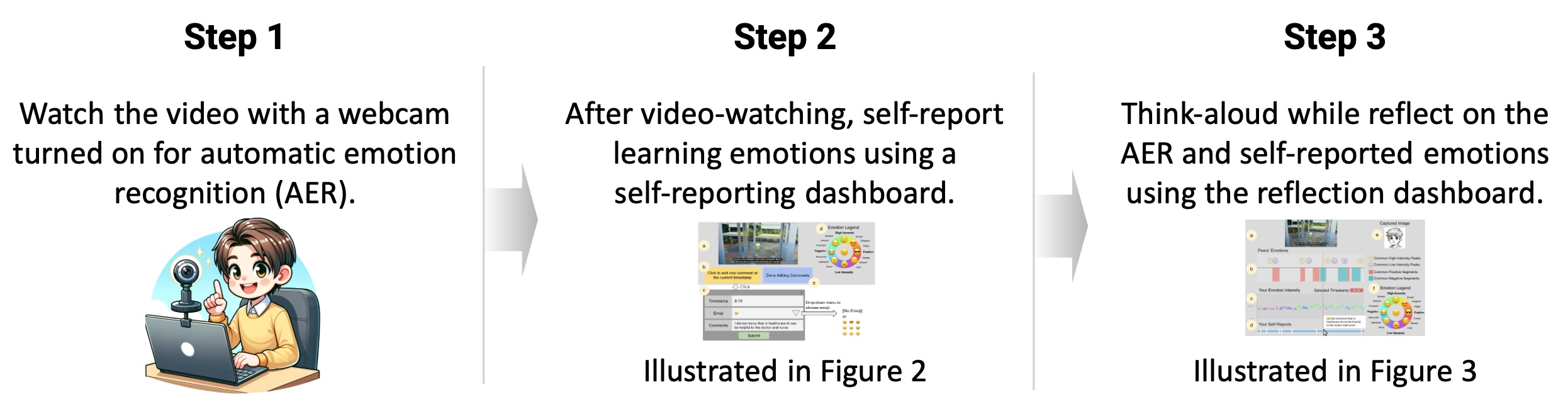}
    \caption{Three major steps of the study process. The self-reporting dashboard and reflections dashboard are illustrated in Figure~\ref{fig:selfrepointerface} and Figure~\ref{fig:flow}, respectively.}
    
    \Description{}
    \label{fig:threesteps}
\end{figure*}

\subsection{Procedure}
We conducted a mixed-method user study remotely through Zoom, an online communication platform.
Upon joining the remote meeting room, we briefly described the study process to each participant, who then read a statement about the study and filled out a consent form.
Participants acknowledged that their facial expressions would be captured on camera during video-watching and processed by AER technologies.
To minimize privacy concerns, participants were informed that video-watching was the only time when their cameras had be turned on. We also acknowledged that captured facial data would not be shared with others and would be destroyed at the end of the user study sessions. Only unidentifiable AER facial data would be stored and analyzed. We explained to the participants prior to the study that their data would be used to understand how AI-enabled learning can be made more inclusive, and the prototype they experienced would be evaluated with other learner populations in similar ways.

Participants in the DHH group were given the choice to use either American Sign Language (ASL) as their sign language or English for communication throughout the study.
Participants who opted for ASL were paired with a team member who was an ASL interpreter or a DHH RA.
Additionally, instructions for the DHH group might differ at specific stages as it was required to provide alternative materials to assist participants in completing the process using ASL.

The learning prototype was inspired from prior work \cite{mirrorus} that captured emotional experiences and provides a reflexive learning process with own and peers' emotions. 
We hosted the prototype in a remote access server on Amazon Web Services.
Participants were asked to open the prototype on a web browser and then share their screens for live observation while interacting with the prototype, with screen recording enabled throughout the study by a researcher. 
Researchers took note of behaviors and emotions observed during the study that would be used to guide the semi-structured interview. The study was conducted in the following steps (Shown in Figure \ref{fig:threesteps}).

\textbf{Step 1: AER During Video-watching}
Participants were first instructed to watch a 15-minute educational video about Augmented Reality with the options of enabling error-free closed captions. Note that all participants showed interests in Augmented Reality technology but only had limited knowledge prior to the study. We chose to use existing AER technology to recognized participants' facial expression during video-based learning process. Specifically, our prototype captured ongoing image streams from cameras, identified participants' facial expressions through MTCNN \cite{deng2020multitask}, and then streamed these images into an AI model (OpenFace \cite{baltruvsaitis2016openface}) to predict valence-arousal value pairs \cite{russell1980circumplex}.
Valence, represented on the horizontal axis, varies from negative (unpleasant emotions) to positive (pleasant emotions).
Arousal, represented on the vertical axis, ranges from low (calmness) to high (excitement). We collected six frames of facial expressions per second.
To note, these valence-arousal pairs were not presented to participants while watching the video, but instead were presented in a dashboard at a later step.



\begin{figure*}[ht]
    \centering  
    \includegraphics[width=0.85\textwidth]{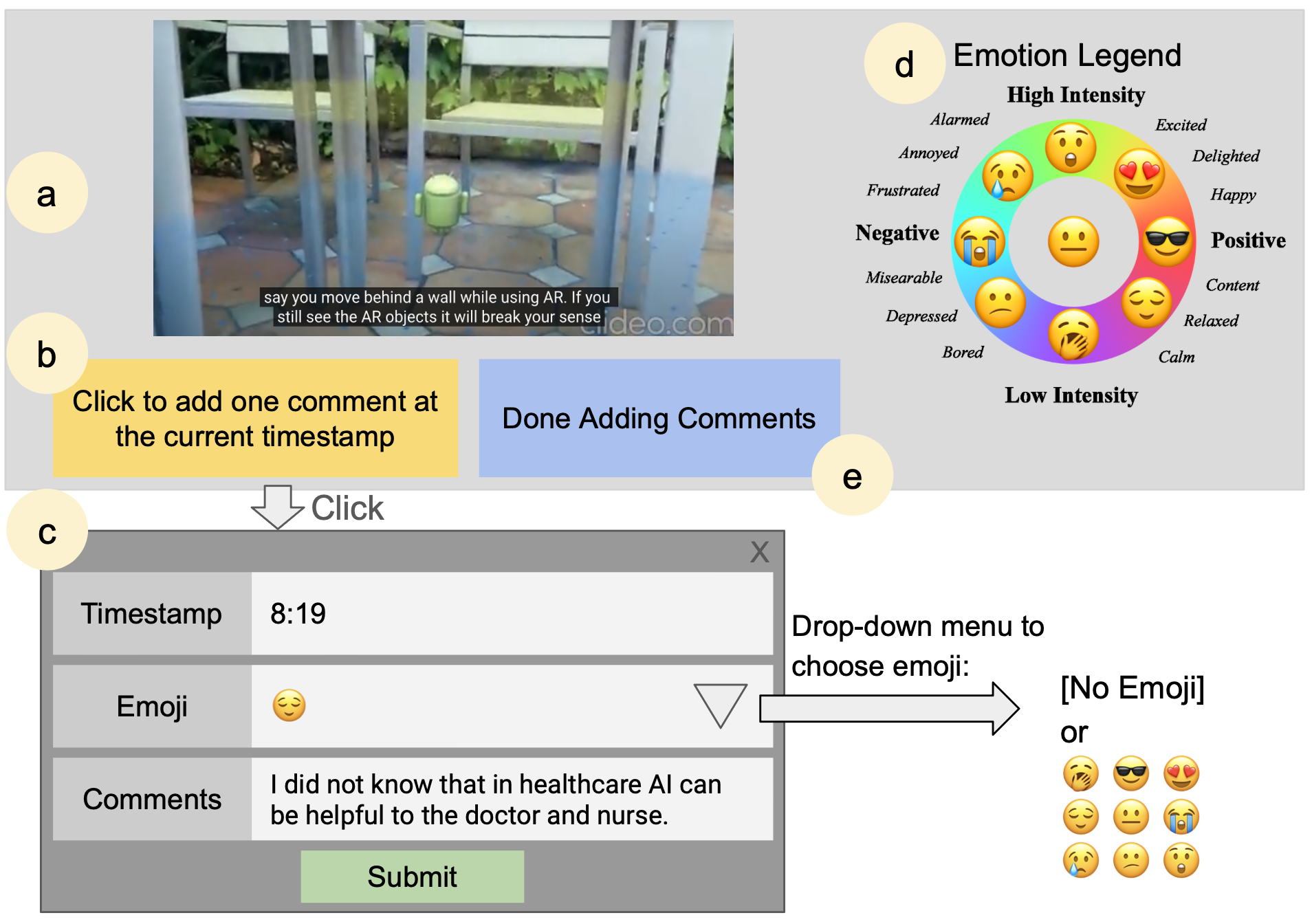}
    \caption{Interface for Step 2, self-reporting emotions in our prototype for video-based learning. This interface contains the following components: (a) A video player that had the same educational video they watched in previous steps. Participants used the video player to choose the timestamp they would like to self-report emotion at. (b) A button to make self-reported emotion at the current timestamp that the video was paused at. (c) A pop-up window after clicking on (b) to make self-reported emotion at. This pop-up window has three parts: the current timestamp that the video was paused at, a drop-down menu where participants may choose from nine emojis or no emoji, and a text comment box for optional additional text comments. (d) Emotion legend  \cite{kucher2018state, hsu2013seeing, chen2022mirror, mirrorus} which participants may refer to. The emotion legend shows the similar categories of self-reported emojis on the arousal-valence axis. (e) A button to continue to the next step using a post-learning dashboard, shown in Figure \ref{fig:flow}.
    }
    \Description{}
    \label{fig:selfrepointerface}
\end{figure*}

\textbf{Step 2: Self-Report Emotion after Video-watching}
After finish watching a 15-minute video, participants were transitioned into a new interface, shown in Fig. \ref{fig:selfrepointerface}, where they were prompted to provide self-reports by selecting timestamps from the progress bar in the video player. Self-reported emotion is important in studying and improving AER, as it serves as a reported emotion that could be used to compare with AER \cite{kaur2022didn, lbw}.
Participants were instructed to provide at least one self-report, and they could add as many self-reports as they wanted.
In each self-report, participants provided their perceived feelings by selecting a emoji and/or writing a text in English at the selected timestamp. Participants would self-report their emotion at a selected timestamp by choosing an emoji from a drop-down menu and type text comments in a pop-up window.   Nine emojis were provide that represented different combinations of affective states, and were selected based on the prior works \cite{sazzad2011affect, reilly2001external}. They may refer to an emotion legend \cite{kucher2018state, hsu2013seeing, chen2022mirror, mirrorus} which shows how each emoji represents on a valence-arousal scale. The text labels in this emotion legend were designed intentionally to NOT align with the emoji; instead, they were only for reference based on prior research. Participants would decide whether they want to use the nearby emoji to represent their desired emotion in text format. For text comments, they were instructed that they may include text comments in their self-reports and were not further instructed what to be included in these text comments. Participants were given the choice to NOT pick any emoji and/or NOT type any text comments.

\begin{figure*}[ht]
    \centering  
    \includegraphics[width=0.85\textwidth]{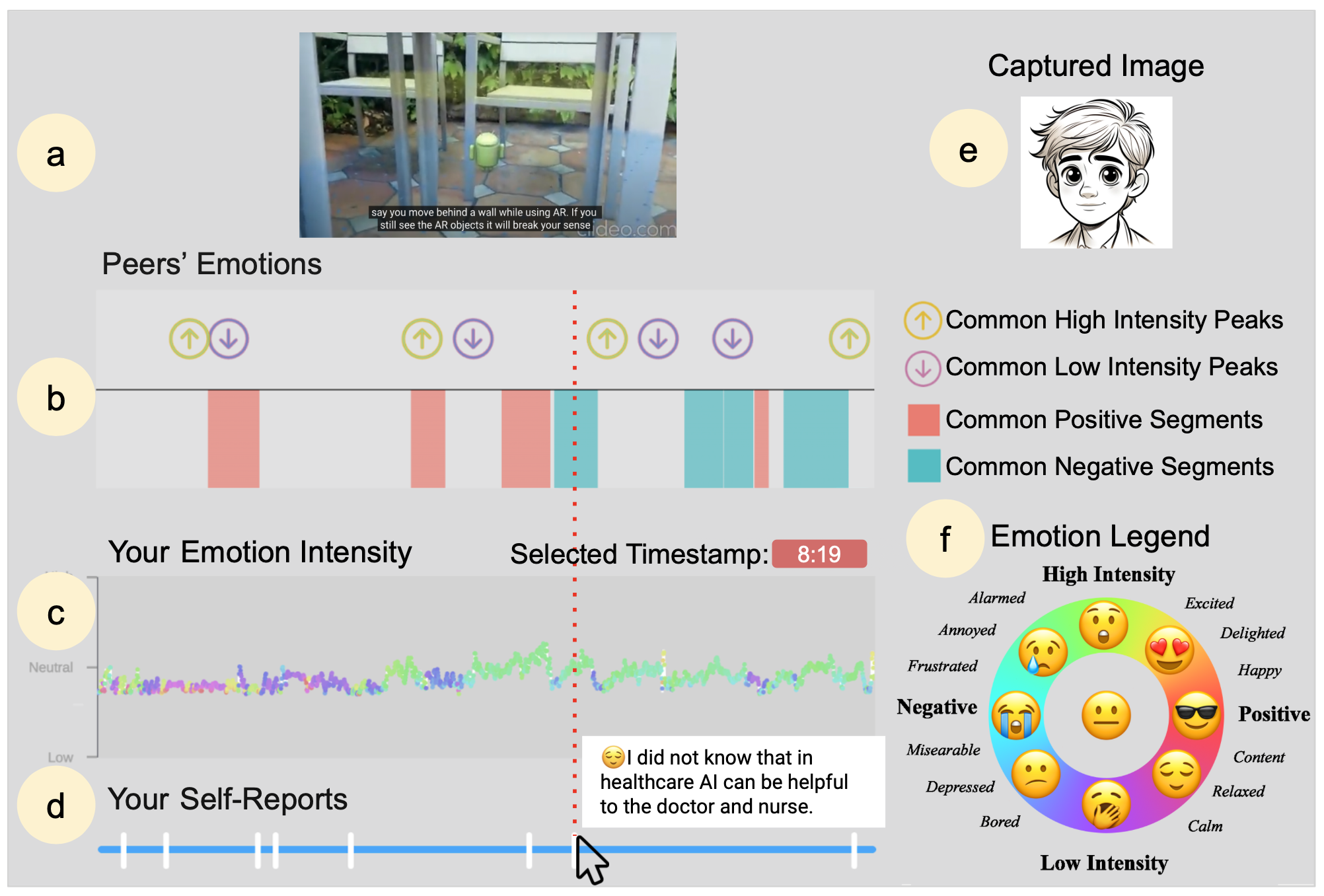}
    \caption{Interface of the post-learning dashboard in our prototype for video-based learning. Participants employed think-aloud protocol while reflecting The dashboard contains the following components: (a) A video player that had the same educational video they watched in previous steps. The video player would fast-forward to any timestamp the participant selected in the components below. (b) Peers' emotions. The peers' emotions are aggregated high \& low emotional spikes (above), and positive \& negative emotions segments (below) from peers. Participants may refer to the legends on the right of each type of peers' emotions. (c) Participants' own emotion intensity line chart over the duration of the video. The line curve represents the high and low intensity (arousal) of participants, whereas the warm and cool colors represent the positive and negative emotions (valence). (d) Participants' self-reports. Participants may over onto a tick to review the emoji and text comments in that self-report. (e) Captured image by AER. The dashboard shows the captured facial expression image at a selected timestamp for participants to review and understand AER outputs. (f) Emotion legend, which is the same as in Figure \ref{fig:selfrepointerface} (d). The emotion legend shows a color-coding of valence for emotion intensity from (c). The sample line graph in (c) is from DP17.}
    \Description{}
    \label{fig:flow}
\end{figure*}

\textbf{Step 3: Think-Aloud while Reflect on Own and Peers' Emotions}
In this step, participants had access to a post-learning dashboard (Shown in Figure \ref{fig:flow}) that allowed them to review and scrutinize their own AER emotions as well as peers' emotions.
Participants were instructed to employ the think-aloud protocol while interacting with the dashboard, which required them to articulate their perceptions and thoughts throughout the entire exploration.
The objective was to reveal their motivations behind each of their actions during the process, with the intention of eliciting these motivations through continuous reflection and articulation. 
For DHH participants, we offered them to choose any way they were comfortable with expressing themselves during think-aloud protocol \cite{DIS2023}, which included, but not limited to, ASL, written English, and spoken English.

During this step, participants were instructed to explore emotions for a time-wise comparison of reflexive learning in two parts: Part 1 focused reflecting on on own emotions, and Part 2 is on reflecting on  peers' emotion. Participants were asked to review and interact with their AER intensity graph of the valence-arousal pairs that were detected during the video-watching.
Participants were told they could click anywhere on the intensity graph because it functioned like a progress bar of the video player.
Upon clicking on the graph with a red line indicating their selection, the video player would fast-forward to the corresponding timestamp to help participants recall video contents, as shown in Figure \ref{fig:flow}. Participants could re-watch the video for any length of time if they wanted. 

Participants could also click on two buttons to check their peers' emotions. Peers' emotion has been found useful to assist learners be more aware of their own emotional experiences \cite{kaur2022didn,mirrorus}. 
There are two types of peers' emotions shown in Figure \ref{fig:flow} (c), following the two approaches presented in prior works \cite{kaur2022didn, mirrorus}: \textit{locative}, marked with up and down arrows representing high and low intensity spikes of peers' emotion at different parts of the video; and \textit{temporal}, marked with red and blue blocks representing positive and negative segments of peers' emotion.
These peers' emotions are self-reported emotion collected from a exploratory study before this study using the same interface used for Step 2 for this study (shown in Figure \ref{fig:selfrepointerface}. Those participants were 16 hearing individuals with the same recruiting criteria, and all of them voluntary agreed to share their unidentifiable self-reported emotions for use in future studies. However, we did not specify the background of these peer students, such as their hearing ability, but rather accommodate each participant's unique education background (e.g., mainstream school vs special education school) to interpret the peer learners they were familiar with. We asked about their understanding of peer learners later in the interview.
Participants could also clicking anywhere on the peers' cues region to fast-forward to the corresponding timestamps in the video, which allowed them to compare their own emotions with those of their peers.

\textbf{Survey}
After exploring with the dashboard, participants were asked to complete an exit survey that was designed to collected their background information, and measure their perceptions when using the learning prototype with a focus on peers' emotions. We chose these questions because in our exploratory studies, DHH learners showed most interests in peers' emotions.
Participants were ask to rate agreement on a seven-point Likert scale (1 - Strongly Disagree, 7 - Strongly Agree). Please see Appendix \ref{appendix: survey} for a list of survey questions.

\textbf{Semi-structured Interview}
Upon completing the exit survey, we proceeded with a semi-structured interview to gain a better understanding of their challenges and perceptions throughout the experience with the reflexive learning prototype.
Sample interview questions included: ``\textit{What part of the video-based learning experience did you like and why,}'' ``\textit{What is the most insightful you notice between the AI-recognized intensity graph, Temporal cues, and Locative cues and why,}'' and ``\textit{Is there any information that you would like the tool to provide, but it is not currently supported.}'' 
When researchers observed behaviors in the previous steps (e.g., re-watch the video while providing self-reports), we questioned participants about their motivation behind these behaviors.



\subsection{Data Analysis}

At the end of each study, we collected the following data: self-reported emotions with emojis and/or text explanations, AER data in valence-arousal pairs along the video timeline, transcripts of think-aloud protocol while reflecting with participants' different types of emotions, survey responses, and interview transcripts. 

\subsubsection{Self-Report Emotions (RQ1)}

For self-reported emotions, we first quantitatively analyzed the number of self-reports made by DHH and hearing participants. We also grouped the emojis based on their representing valence and arousal categories, respectively, based on prior works \cite{sazzad2011affect, reilly2001external}. Each group contains three emojis. Specifically, we grouped the emojis into positive, neutral, and negative for valence (three columns from right to left in emotion legend shown in Figure \ref{fig:selfrepointerface} (d), as well as high, neutral, and low intensities for arousal (three rows from top to down in emotion legend shown in Figure \ref{fig:selfrepointerface} (d)). The number of emojis reported by DHH and hearing participants was compared using statistical analysis. We also examined the text comments for self-reports without an emoji.


\subsubsection{Reflect on AER and Self-Reported Emotions (RQ2)}
For each sentences in the think-aloud step, we transcribed the video recordings to analyze how participants perceived the two different types of emotion data together on our dashboard (RQ2). For the DHH group, the think-aloud protocol completed in ASL was transcribed either by the hearing member who was an ASL interpreter or a DHH RA. For the hearing group, two hearing team members reviewed the video recordings and transcribed the spoken sentences into plain English text. We performed open coding on think-aloud transcriptions to establish initial themes. Then, we performed axial coding based on initial themes using the frameworks of meta-cognition and social meta-cognition developed by \cite{chiu2009social} and meta-cognition for SRL developed by \cite{zimmerman2009self}.
Two hearing team members independently read and coded 25\% of the transcripts, and followed by multiple rounds of discussion to compile their codes until a consensus was reached.
The remaining 75\% of data was then coded individually by one hearing team member.

\subsubsection{AER (RQ2)}
Statistical analyses were performed on the participants' AER-detected emotions (i.e. valence-arousal pairs based on facial recognition) during video-watching to gain insights into the emotional experiences developed between DHH and hearing participants.
The overall distribution of AER data between DHH and hearing participants was first compared and no differences were found. Second, informed by DHH participants' interest in peers' emotions during the interview study, we compared the mean value differences of the emotions associated with the talking-heads and non-talking-heads \cite{guo2014video} in the educational video between the two groups were examined. We reviewed the entire educational video from start to end to manually label the talking-head intervals. Each talking-head interval had a start and end time (e.g., 1:02-1:30). Those interval sections that were not labeled were considered non-talking-head intervals. Note that we analyzed arousal and valence values separately, which was also performed in prior works (e.g., \cite{alluri2015musical, delatorre2019impact}).



\subsubsection{Survey (RQ2)}
The survey contained questions to understand participants' perception toward ethical concerns with AER technology through their perceptions with peers' emotions (RQ2). For each question, the Mann-Whitney U test was used to analyze the difference in responses between DHH and hearing participants for each questions. 

\subsubsection{Interviews (RQ1 and RQ2)}
For interview transcripts, we first transcribed the recordings. For DHH participants who opted to complete interview with ASL, the study recordings were transcribed either by the hearing researcher who is ASL interpreter or by a DHH researcher. For other participants, the recordings were auto-captioned through Zoom recordings. All recordings that contained identifiable information were deleted after transcription was completed.
We conducted an inductive thematic analysis \cite{braun2006using} to analyze participants' interview feedback. Two hearing and two DHH RAs independently open-coded five transcripts from DHH and hearing participants respectively, which formed the basis for the initial codebook. The remaining transcripts from DHH participants were coded by one of the RAs after the initial codebook was established.
Following several rounds of discussion and analysis of these transcripts, the team iteratively refined the codebook into a structured organization of codes into coherent themes, focusing on the divergent experiences of DHH learners. One hearing team member then completed the analysis of remaining transcripts with hearing participants. Major themes included: \textit{ethical concern on technology}, \textit{technology related behavior}, \textit{suggestions to meet emotional needs}. 


\section{Findings}



\subsection{Usage and Perception of Self-Report Emotions (RQ1)}\label{sec:rq2finding} 
For RQ1, we analyzed DHH and hearing participants' self-reported emotion and their interview data.

\subsubsection{DHH Participants' Different Emoji Interpretation and Usage} 
Overall, 20 DHH participants made 157 self-reports (M=7.64, SD=3.39), of which 48 (30.57 \%) self-reports did NOT contain an emoji, whereas 20 hearing participants made 79 self-reports (M=3.95, SD=2.16), of which 14 (17.78 \%) self-reports did NOT contained an emoji. 
For DHH group, nine (45 \%) participants did not choose emoji in at least one of their self-reports, including six (30 \%) participants who did not choose an emoji in more than two-thirds of their self-reports. For hearing group, four (20 \%) participants did not choose emoji in any of their self-reports, while the other 16 participants used emoji in all their self-reports.

However, the text of these self-reports without emojis from DHH group still included some words describing emotion such as ``interesting,'' ``funny,'' and ``curious.'' By comparing the length of text between self-reports with and without emojis using a two-sample t-test, self-reports without emojis (M=26.19, SD=27.63) from DHH participants were significantly longer than those with emojis (M=10.11, SD=13.45) (t(57)=3.83, p<.001). 
This indicates that despite not choosing an emoji, DHH participants were still willing and able to express their emotions through text. 
During the interview, 12 out of 20 (60 \%) DHH participants mentioned that they found emojis unable to accurately express their emotions. For example, D11, who did not choose an emoji in all their 10 self-reports, explained during the interview, ``\textit{Because I could not put my real feelings into emoji. I felt that it is easier for me to typing than using emoji.}''


\textbf{DHH Participant's Design Idea: Alternative Methods for Expressive Communication}
DHH participants also found the current selection of emojis inadequate to represent all their emotions. For instance, D3 found it hard to express their confusion with the current selection, ``\textit{I think adding the emoji with the raised eyebrow would be better for showing confusion. That would be a good one to add: Confused... I think it’d be better to add an emoji that matches the confusion/don’t understand emotion, maybe.}'' Eyebrows were considered an important facial movement in ASL and Deaf culture to convey emotion, while the participant commented that most emojis available online, beyond our current emoji selection, do not have eyebrows, making it hard to select emojis that accurately represent their emotions.

During interviews, DHH participants suggested that instead of showing peers' emotions visualization after learning, they would prefer to see the emotions in more expressive formats. This would better assist them to be aware of and reflect on their own emotions with the video content. They also suggested that the peers' emotional cues could be ASL comments rather than emojis and could pop up during the video and they do not have to stop the video to read them.  For example, D16 mentioned that the peers' sign language comments could be presented ``\textit{live beneath the video... we could set up an area on the screen to allow participants to upload their sign self-reports, like a pop-up.}'' 

\subsubsection{DHH Participants Enhancing Emotional Awareness Through Rewatching} 
During self-reporting, DHH participants on average took a longer time compared to hearing participants. Therefore, each participant's system action log during the self-report step was analyzed. An annotation was made for each action the participant performed with the video player, including play, pause, rewind forward, and rewind backward. We found that DHH and hearing participants demonstrated different strategies for making self-reports. 

The action ``play video'' was initiated at least once by 14 (70 \%) DHH participants compared to seven (35 \%) hearing participants. Eight (40 \%) DHH participants re-watched the whole 15-minute video from the beginning to the end during self-report, whereas none of the hearing participants watched the whole video. Instead, hearing participants would use the preview window when scrolling through the progress bar on the video player and seek forward or backward to the desired timestamp to make self-report. By re-watching while self-reporting, results of a two-sample t-test showed that the DHH group made significantly more self-reports compared to the hearing group (t(38)=4.25, p<.001).


During the interviews, DHH participants explained they enjoyed self-reporting and sharing their emotions potentially but lacked the confidence to do so without rewatching. Because they were worried that they missed information about the video which made it misleading for themselves. D17, who re-watched the whole video during self-report, explained, ``\textit{Information missed on the first watch can be gained on the second watch. You'll notice a little bit of a small difference when you watch it a second time. This will help with memory.}'' Another related reason for rewatching was that the captioned video was ``too flat,'' reading the captions felt forced, creating a feeling of temporal pressure and making them less aware of their emotions. D4 noted, ``\textit{reading captions can feel a bit lackluster, which makes it difficult to know if the speaker is using high or low intonation. Of course, we can use facial affect for clues, but the captioning itself is still flat.}'' Rewatching allowed them to feel more confident, get to pay attention to more details beyond captions, and be more relaxed and aware of emotions.




\textbf{DHH Participant's Design Idea: Lightweight Initial Markers for Later Expansion} Participants suggested improving the self-reporting emotional experience to alleviate recalling pressure. They felt reporting emotions while watching the video would disrupt learning and agreed on the necessity of understanding the video's key idea before describing their emotions. Emojis could serve as time anchors for later reference, but details should be expanded after grasping the video's key idea. As D3 noted, ``\textit{I prefer describing emotions after fully remembering, to keep watching and self-report separate.}'' For example, D3 suggested marking emojis during the video and adding text later, ``\textit{Clicking an emoji in the moment shows true feelings, which can be expanded on later.}'' D12 suggested allowing participants to self-report emotions while watching the video, helping them ``\textit{remember how the content made them feel and the related information}.''

\hspace{0mm}

\textbf{Summary (RQ1)} 
DHH participants found emojis inadequate to accurately reflect their emotions. Compared to hearing participants, DHH participants opted for longer text descriptions significantly more often despite that written English is not the first language of many of them. They suggested more expressive alternatives, such as ASL comments or improved emoji designs, to better convey their feelings. Rewatching the video allowed DHH participants to self-report emotions more confidently and in detail, as they felt the video captions were ``too flat'' and other visual details couldn't be fully understood without rewatching.


\subsection{Usage and Perceptions of AER-Generated Emotions (RQ2)}\label{sec:rq1finding}

For RQ2, we mainly analyzed DHH and hearing participants' think-aloud, survey, and interview data.

\subsubsection{DHH Expressed Fewer Reflection Sentences, Yet Prioritize Peers' Emotion Cues} \label{peers}

\begin{figure*}[ht]
    \centering  
    \includegraphics[width=0.8\textwidth]{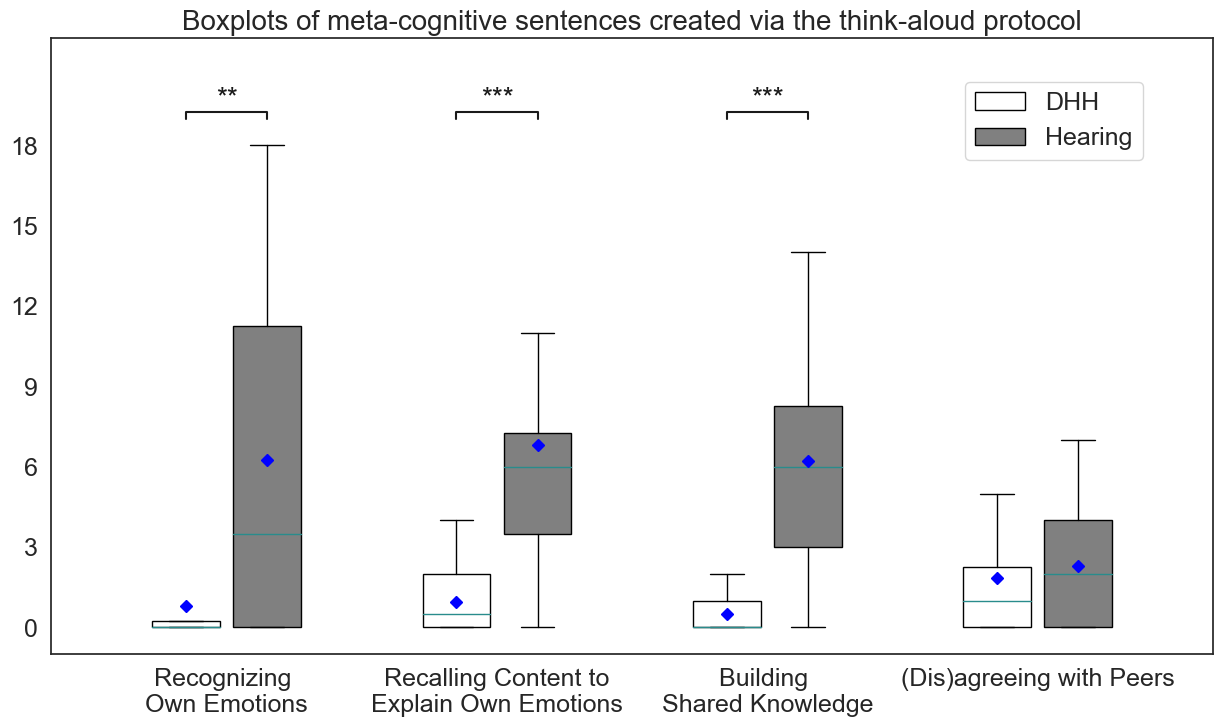}    
    \caption{Boxplots of the four types of meta-cognitive sentences created via the think-aloud protocol between the DHH and hearing participants. Mann-Whitney U-tests suggested that hearing participants expressed significantly more sentences on \textit{Recognizing Own Emotions}, \textit{Recalling Content to Explain Own Emotions}, and \textit{Building Shared Knowledge} than DHH participants. ** denotes a significant difference with p<.01, and *** denotes a significant difference with p<.001.}
    \Description{}
    \label{fig:think-aloud}
\end{figure*}

During the think-aloud phase, participants in both groups reflected on their own and their peers' emotions using the learning analytics dashboard.
As a result of their reflection sentences, including the transcription from ASL, four types of meta-cognitive processes were identified. For each meta-cognitive process, the number of sentences between DHH and hearing participants was compared using Mann-Whitney U tests. Figure \ref{fig:think-aloud} shows the boxplots of each of the four types of meta-cognitive processes for DHH and hearing participants. 

The first two meta-cognitive processes are related to participants' own emotions.

\begin{itemize}
    \item \textit{Recognizing Own Emotions.} Participants recalled or described their own emotions without recalling and retrieved video content (e.g. \textit{And this time, I feel maybe a little bored}, \textit{This graph shows I am pretty positive}). 
    DHH participants expressed 16 sentences (M=0.8, SD=1.54), whereas hearing participants expressed 125 sentences (M=6.25, SD=6.38). The Mann-Whitney U-test showed that hearing participants expressed significantly more sentences in ``Recognizing Own Emotions'' compared to DHH participants (U=91.5, p<.01).
    \item \textit{Recalling Content to Explain Own Emotions.} Participants recalled and retrieved video content to explain their own emotions (e.g. \textit{For here, I see the Augmented Reality, I feel very excited about it}). There were 19 sentences from the DHH participants (M=0.95, SD=1.19) and 136 sentences from the hearing participants (M=6.8, SD=5.75). Hearing participants expressed significantly more sentences in ``Recalling Content to Explain Own Emotions'' compared to DHH participants (U=43.5, p<.001). 
\end{itemize}

There are two other meta-cognitive processes related to peers' emotion.

\begin{itemize}
    
    \item \textit{Building Shared Knowledge.} Participants explained how and why they (dis)agreed with and/or (fail to) understand peers’ emotions in detail (e.g. \textit{I don't agree I got very excited by it, but I was definitely feeling more on a positive side}). DHH participants expressed 10 sentences (M=0.5, SD=0.76), whereas hearing participants expressed 124 sentences (M=6.2, SD=4.80). Hearing participants expressed significantly more sentences with ``Building Shared Knowledge'' compared to DHH participants (U=30.0, p<.001).
    \item \textit{(Dis)agreeing with Peers.} Participants only encapsulated whether they (dis)agreed with and/or understood the peers’ emotions without further explanations (e.g. \textit{Yeah, I do agree}). DHH participants expressed 37 sentences (M=1.85, SD=2.48), while hearing participants expressed 46 sentences (M=2.3, SD=2.15). There is no significant difference between hearing and DHH participants (U=166.5, p=0.36).

\end{itemize}

\subsubsection{DHH Participants' Challenges in Describing Visualizations Due to Language Diversity} 

During the think-aloud sessions, DHH participants frequently encountered challenges when describing visualizations on the dashboard interface due to language diversity. In the think-aloud sessions, the most common communication mode for DHH participants was ASL. Most of the Deaf participants exclusively used ASL to express their thoughts during the sessions, whereas HoH participants displayed a mix of communication modes, including spoken English, ASL, and sometimes a combination of both \footnote{ASL and spoken English are two different language that have their own unique grammatical structures. ASL is commonly used in the DHH community in the US.}. This varied use of communication modes often involved blending ASL with spoken English or switching between the two depending on the context and the participant's background.

Most participants did not use the explicit labels and terminology provided on the interface, opting instead for a variety of visual-gestural movements (which are not signs in ASL) that embodiedly describe visualization. For example, D7 used hand movement pointing upward to refer to the intensity spike in her line graph, as opposed to fingerspelling \footnote{Fingerspelling is the process of spelling out words by using hand shapes that correspond to the letters of the word. 
If an ASL/English bilingual person wishes to express a concept for which they know the English word but for which there is no existing sign and there is no convenient method of combining other signs to express it, and the signer wants to specify a less common meaning of that sign -- then there is a high probability the person will fingerspell it.} the term ``SPIKE'' or using the specific terminology presented on the interface (Figure \ref{fig:flow}(c)). Similarly, D9 traced the shape of the line graph in the air with their finger to visually describe the peaks and valleys of their emotional intensity timeline, smaller zig-zag for segments that has smaller fluctuation, and bigger zig-zag for segments that had bigger fluctuation. While this method effectively conveyed the shape of the graph. 

This resulted in ambiguous references that required researchers to ask clarifying questions to understand which part of the interface was being discussed. Furthermore, while ASL signs such as "UP" and "DOWN" were used by multiple participants, their meanings varied depending on the context, complicating the researchers' ability to accurately interpret the participants' feedback. Some referred to Figure \ref{fig:flow} (b) arrows, while others referred to Figure \ref{fig:flow}(c) peaks. 

Interview data explains that the reliance on visual-gestural expressive mode rather than explicit terminology either in fingerspelling or spoken English was because the participants were unfamiliar with the specific terms used on the interface, did not have an exact ASL sign for it, or found the fingerspelling less intuitive for communication. This issue highlights a significant barrier of language diversity in describing visualizations and fully demonstrating their ability to reflect on their learning experience, which explains why the quantity of three out of four meta-cognition processes is significantly lower for DHH participants compared to hearing participants.

\textbf{DHH Participants' Design Idea: ASL Comments from Peers to Enhance Reflection} The interview data suggested that DHH participants had a strong preference for seeing their peers' direct reactions as a means to initiate conversations, rather than solely reflecting on their own experiences. Participants frequently struggled with the asynchronicity between signing and interacting with the dashboard. Visual indicators of peers' emotional responses, such as intensity spikes on a graph, were not sufficiently intuitive or perceived as connected to conducting reflection. 

Participants proposed using real-time peers' ASL comments to support immediate discussions during and after learning activities. For instance, when a peer's graph showed a positive emotion spike, participants suggested signing "yes/no" to start a conversation about their own feelings and would feel more natural elaborating on that if it were with another signing person. ASL, being a more native language for many participants, would facilitate these conversations, allowing them to feel more connected and comfortable when discussing their emotions. This approach would help bridge the communication gap that often exists in virtual environments. D3 commented that seeing peers' emotions made them feel the authenticity of peers' existence \textit{``I liked seeing those perspectives from my peers even if they differed from mine because everyone has varying perspectives. That shows their authenticity.''}

\subsubsection{DHH Participants' Trust in AER More Fluctuated by Peers' Cues}

The post-study survey showed that both hearing and DHH participants agreed that peers' cues increase their privacy concerns (DHH: Mean=4.8, SD=2.1; hearing: Mean=4.2, SD=1.9), help them notice the limitations (DHH: Mean=5.4, SD=2.0; hearing: Mean=4.8, SD=2.3) and actual evidence (DHH: Mean=5.2, SD=2.1; hearing: Mean=5.2, SD=1.8) of tool, with no significant differences between DHH and hearing groups using Mann-Whitney U test. 

As for the two trust-related questions, significant differences were found between DHH and hearing group - DHH participants felt more increase and decrease in trust. In detail, DHH participants on average rated agree (Mean=6.0, SD=2.8) while hearing participants on average rated moderately agree (Mean=5.0, SD=1.9) for question \textit{Navigating peers' emotions/cues increases my trust in the tool than only seeing my own emotion} (p<.05). DHH participants on average rated moderately agree (Mean=5.2, SD=2.1) while hearing participants on average rated agree neutral (Mean=3.3, SD=1.1) for question \textit{Navigating peers' emotions/cues decreases my trust in the tool than only seeing my own emotion} (p<.001).

During the interviews, DHH participants explained factors leading to increased and decreased trust toward AER technology. They mentioned that knowing peer learners also used the same tool, as well as their perception that DHH individuals would usually express their emotion via facial expression more ``\textit{honestly}'' (D3) could increase their trust toward peers' AER emotions. Meanwhile, they showed concerns about the accuracy of recognizing DHH individual's facial expressions with AER technology, which led to a decrease in their trust toward peers' AER emotions. The hearing participants, on the other hand, either did not mention such concern or only briefly mentioned it without giving detailed explanations. 
While the background information of peer learners was not revealed to our participants, DHH participants perceived their peer learners to be DHH individuals as well.

Among DHH participants, a primary concern was the unnatural feel of the expression capture process compared to more spontaneous signing conversations among DHH learners. This setup often led to nervousness, which participants felt compromised the naturalness of their facial expressions. This nervousness, in turn, could lead to inaccuracies in the results, potentially increasing their anxiety about being under \textit{scrutiny} by the AER system. For example, D20 found themselves ``\textit{scared by the fact that this tool recognizes my face.}'' They further explained, ``\textit{when people are afraid, they often don't show their real reactions, and how can AI identify this ... If you tell a deaf person that the AI will recognize his face, they will become alert ... they will act differently than they really are.}'' Additionally, DHH participants felt unsure about using current AER technology as it might recognize their Deaf identity based on AER emotions and possibly leak their identity. For example, D18 said, ``\textit{I would feel insecure if the AI shares my information with the outside world(Deaf community) and that the AI is stealing my identity.}'' 

No design ideas were given on AER and trust. 


\subsubsection{DHH Participants' Curiosity Towards Similar Emotion Patterns with Peer Learners}
\label{talking-head}
In the interview, DHH participants expressed interest in seeing other emotion patterns and if they shared the same emotions. Unlike hearing participants, who had more varied interests in emotion patterns, seven DHH participants specifically wanted to see if their peers shared their complaints about the visual presentation of the video, especially segments with "talking-heads." We found some self-reported emotions within those segments (e.g., D3 wrote, ``\textit{The woman appears a bit stone-faced. It was boring to watch with such a ‘bland,’ blank face. She didn’t look excited about the AR project.}'') reflecting this sentiment. However, self-reported emotion cannot provide continuous emotion data to answer DHH participants' questions. Therefore, following the participants' curiosity, we analyzed DHH participants' AER. To investigate whether AER can capture the changes, the video was categorized into two types based on prior works \cite{guo2014video}: talking-head and non-talking-head. 



\begin{figure*}[ht]
    \centering  
    \includegraphics[width=1.0\textwidth]{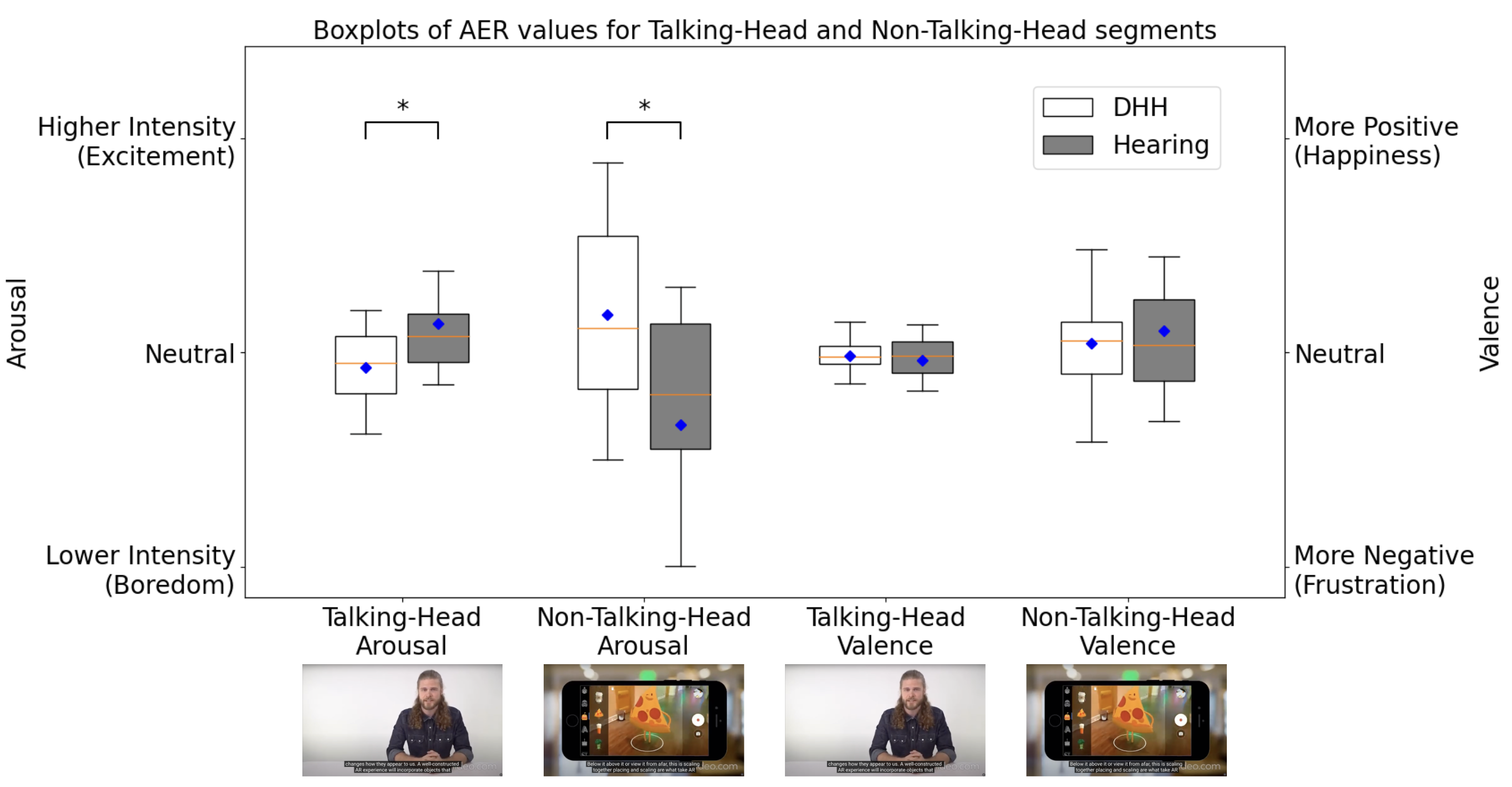}    
    \caption{Boxplot of valence and arousal values for talking-head and non-talking-head segments for DHH and hearing participants. The 18 talking-head segments and 17 non-talking-head segments appeared alternatively throughout the video. Two-sample t-tests showed that there are significant differences in arousal values for both talking-head and non-talking-head segments between DHH and hearing participants. Specifically, DHH participants showed lower intensity during talking-head segments and higher intensity during non-talking-head segments. * denotes a significant difference with p<.05.
    }
    \Description{}
    \label{fig:talkinghead}
\end{figure*}

In total, there were 18 segments of talking-head totaling 646 seconds (M=35.89 seconds, SD=21.24) and 17 segments of non-talking-head totaling 255 seconds (M=15.0, SD=7.03). The talking-head and non-talking-head segments were presented alternatively in the video (e.g., a non-talking-head segment between two talking-head segments). Figure \ref{fig:talkinghead} includes sample screenshots of talking-head and non-talking-head in the video. For each type of video, the mean and standard deviation of arousal and valence values was calculated for each participant. Since each participant's common facial expression differed, the mean values of valence and arousal for each type of video were further adjusted by deducting the ``baseline'' emotion for each participant, that is, mean values of valence and arousal for the entire video, respectively.

Two-sample t-tests were used to compare the adjusted arousal and valence means for talking-head and non-talking-head segments between DHH and hearing participants. The results showed that for talking-head segments, DHH participants had significantly lower intensity (arousal values) compared to hearing participants (t(38)=-2.40, p<.05). In contrast, for non-talking-head segments, DHH participants had significantly higher intensity compared to hearing participants (t(38)=2.47, p<.05).
There are no significant changes in valence dimension between DHH and hearing participants for either talking-head or non-talking-head segments. Figure \ref{fig:talkinghead} shows the boxplots of valence and arousal values for talking-head and non-talking-head segments for hearing and DHH participants.
Four repeated measures ANOVAs were also used to compare the effect of talking-head vs. non-talking-head style on the valence changes (-1 to 1 representing negative to positive, with 0 meaning neutral) and arousal changes (-1 to 1 representing low intensity to high intensity, with 0 meaning neutral). To account for individual differences among participants, ``1/PID'' was included as a random effect in the model. The results of repeated measures ANOVAs confirmed the results from two-sample t-tests. In other words, DHH participants felt less excited during talking-head parts of the video and more excited during other parts compared to hearing participants.

\textbf{DHH Participants' Call for Improved Video Presentation Styles} DHH participants expressed a strong desire for better video presentation styles to enhance their learning experience, in addition to their preferences on ``talking-head'' segments. First, participants emphasized the need for videos to clearly outline the subject and highlight key elements. They recommended adding detailed timelines, explanations of vocabulary terms, and more structured content.
Additionally, participants called for enriching the video content by incorporating more engaging visuals and diverse visual components. They suggested adding images and animations to capture viewers' attention and make the content more interesting. Moreover, participants suggested eliminating extraneous or distracting elements from the videos. They felt that the end of the videos often contained too much information, leading to overwhelm and a lack of engagement.

\hspace{0pt}

\textbf{Summary (RQ2)} DHH participants expressed significantly fewer sentences about their own emotions compared to hearing participants due to language diversity challenges but showed similar engagement when responding to peers' emotions. They proposed using real-time ASL comments from peers to support reflection, emphasizing the need for a more intuitive and connected experience. DHH participants' trust in AER was significantly more impacted by peers' ideas. DHH participants wanted to see their peers' emotional patterns, especially reactions towards ``talking-head.'' AER is found to be more supportive of sharing peer learners' emotional patterns, whereas self-reported emotions are too scattered to do so.

\section{Discussions}


Our research provides new insights into DHH learners' emotional awareness in video-based learning by examining AER and self-reported emotion data. We found that DHH learners did not benefit as much as hearing learners due to language diversity and provided suggestions to improve self-report and AER design to be more DHH-centered (RQ1, RQ2).
Below, we first discuss lessons learned from designing emotion technologies with DHH individuals to inform further research that supports other underserved populations. Then we discuss the design implications of making video-based learning more accessible for DHH learners. Finally, we discuss the theoretical implications of SRL on DHH learners.

\subsection{New Research towards Inclusive  Learning Supported by Emotion Technologies} 





\subsubsection{Culturally and Linguistically Aligned Design for Emotion Reception and Expression} In RQ1, an important finding was that the current emoji design was not expressive enough for our DHH participants, leading them to opt for typing text comments, which is often not their first language. Emoji design was connected with ASL language usage and was reported to be not representative enough due to the lack of eyebrows. Additionally, ASL comments were mentioned across both RQ1 and RQ2. Participants suggested providing their own comments in ASL and viewing peers' comments to enhance their awareness of their own emotions and feel more comfortable discussing emotions. This suggests that future research should consider designing emotion reception and expression methods that align with the culture of specific communities.

\subsubsection{Unique Insights from Self-Reported Emotions Process} Our study showed that while AER data did not differ significantly between DHH and hearing individuals (without taking into consideration the interview insight from section \ref{talking-head}), self-reported emotions process and outcome did (RQ1). This suggests that self-reports provide unique insights not captured by AER. DHH learners struggled to recall their emotions due to low confidence in visual content, revealing hidden emotional efforts and potential differences in self-report data. Prior research talked about underserved communities requiring additional emotional efforts in work \cite{hersh2023comparative, johannsdottir2021s}, and we confirm it also existed in providing self-reported emotion. Future research should design different processes to help various learners provide confident self-reports and reconsider how to process self-report data differently.

\subsubsection{Understanding Psychological Impact of AER on Emotion Expression, Vice Versa} While DHH participants considered their facial expressions could accurately present themselves, they also mentioned that being aware of AER technology could impact their natural facial expressions (RQ2), such as being more nervous or scared by the technology, which leads to even lower accuracy of AER. 
Additionally, DHH participants' trust in AER was more influenced by peers' opinions (RQ2), more specifically how DHH individuals perceive the community to view the technology.  It's essential to explore the psychological and social factors that influence how emotion technologies encourage and impede emotional expression in underserved populations, as this might differ from the general population. Additionally, we should consider how these factors impact AER performance in underserved populations.






\subsubsection{Leveraging Different Emotion Data for Learnersourcing} DHH participants were interested in seeing if peers' emotion patterns aligned with their own (RQ2), specifically the ``talking-head'', offering unique perspectives that helped us better analyze AER data and identify learning challenges. This suggests that DHH learners' emotion data should be considered crucial input for learnersourcing, a form of crowdsourcing where learners contribute content for future learners while engaging in meaningful learning experiences. The two types of emotion data, self-reported emotions, and AER, should be better explored together to (automatically) generate meaningful peer cues and related information, creating a more inclusive learning environment for DHH learners. Additionally, our findings may inform other students in underserved communities, considering their unique community characteristics, values, and culture, to rethink how their learning data, including emotion data, could be better utilized for community collective action and benefits.

\subsection{Design Implications for Video-based Learning for DHH Learners}

\subsubsection{Video Segmentation for Effective Re-watch and Recall}
In our RQ1 findings, we found that DHH learners benefited from re-watching the video to self-report their emotions.
Re-watch behavior is an essential factor for hidden emotional effort to bolster confidence. Aside from DHH learners, Kim et al. \cite{kim2023older} found that older adults also benefited from re-watched videos to recall the contents during online learning.
Instead of re-watching the video from the beginning, we suggest enhancing confidence through the video player design by providing segmentation by topic, as well as engagement level using emotion data, and labeling it in the progress bar.
Doing so will not only allow learners with the greater control over where to re-watch \cite{fiorella2018works}, but also decreasing cognitive load \cite{cheng2014investigating}, improving information processing \cite{kruger2016measuring}, and making complex information in discernible structure \cite{biard2018effects, spanjers2012explaining}.
These features can achieve similar functionality like how hearing participants locate specific content in thumbnail previewing.
Besides segmenting by topic, the video can also be divided into smaller chunks of equal length (e.g. 30 seconds each).
Shorter video segments allow learners to periodically assess their own confidence of their video comprehension and determine whether re-watching is necessary during video-watching.
Such video segmentation design can be performed by an instructor or instructional designer, or it may even be processed automatically by emotional technology.

\subsubsection{Video Representation Improvement for Higher Learning Engagement}

In our RQ2 findings, our DHH participants showed more engagement with certain segments of the video, such as ``non-talking-head'' segments. They also discussed about suggestions toward better video representation, such as eliminating extraneous or distracting elements from the video. These visual cues have been found to support learning of scientific concepts that enhance learning time and efficiency \cite{lin2011using}, and visual transitions in the video, e.g., a slide view to a classroom view, are associated with more views on \cite{kim2014understanding}.
Our participants suggested that the educational video they watched could be improved by incorporating more visual examples (such as visual examples of Pokémon Go). Seeing an example that resonate one's memories may tend to evoke stronger emotional responses \cite{dudzik2020exploring, dudzik2020investigating}.
The non-talking-head segments also led to more engagement with the video content for DHH participants. In contrast, the talking-head segments should be reduced as participants felt bored watching a stone-faced speaker.
Therefore, when the speaker has not shown any emotional changes on their face for a period of time, the speaker's talking-head and other visual content could be presented in alternative ways. We suggest future research to work with DHH learners to better understand their preferences in visual representation and propose a taxonomy or guideline for DHH-learner friendly educational videos. Additionally, future research should investigate whether the content of the video may influence the preferred video representation for DHH learners.

\subsubsection{Language-diverse Video Self-Reports for Enhanced Social-Emotional Support}

In our RQ1 findings, our DHH participants suggested alternative methods to express their emotions. In RQ2 findings, they showed interests in reviewing peer learners' emotion to support their own emotional awareness. Our findings suggested that DHH learners could benefit from shared peer learners' emotions other learners studied in prior works \cite{mirrorus, bazarova2015social, semsioglu2018emotionscape}. However, their diverse language background prevented them from fully engage with sharing and perceiving peers' emotions in text and emoji formats. While prior work suggested that DHH learners found it easier to express and perceive ASL-based self-reports \cite{chen2024towards}, such self-reports were hard for other learners who do not use ASL to understand.
Future research could explore methods to bridge the diverse language gap between DHH and hearing learners' self-reports. For example, emotion technology could be used to translate emotional expressions between written language and sign language, thereby supporting language diversity.


\subsection{SRL for DHH Learners} 


SRL is an iterative process involving multiple learning phases, such as performance, self-reflection, and forethought \cite{zimmerman2009self}. In the UDL framework, self-regulation and emotions like curiosity fall under Engagement. We address the gap highlighted by Brandt and Szarkowski \cite{brandt2022universal} regarding limited UDL research for DHH learners. We found differences between DHH learners and hearing learners in the self-reflection phase via think-aloud protocol, measured using meta-cognition. 

Among the various types of meta-cognitions, DHH participants showed minimal engagement in the first three types of meta-cognitions related to describing their own emotions where interpreting emotion graphs with our prototype is necessary (RQ2). This challenge may stem from the limited range of emotion and visualization signs in ASL, restricting their reflective opportunities. Our findings extend Morrison et al.'s \cite{morrison2013deaf} discovery of varied meta-cognitions between DHH and hearing university learners during reading and classroom instruction to online settings, highlighting how technology, especially visualization dashboards, further exaggerates the issue. Equal access to information in learning analytics dashboards does not equate to equal SRL opportunities. The limited STEM vocabulary in ASL may hinder DHH individuals from effectively discussing and benefiting from emerging AI technologies. It is important to consider how language diversity and social-cultural factors influence how AI can be explained and understood by underserved groups. 

Additionally, DHH learners were more likely to reflect on peers' emotions, with their trust in AER influenced by peers' opinions (RQ2). This shows that peers' emotions and ideas play a more important role for DHH learners than for hearing learners, and peers' ideas may become even more crucial when AI is involved. Previous research has highlighted the importance of social transparency in explaining AI to users to calibrate trust \cite{ehsan2021expanding}. Our findings suggest that peers' cues might be even more significant for DHH learners to effectively reflect with technology.

\subsection{Limitations}

We acknowledge a few limitations in our study. First, we only studied with DHH learners as one population with special education needs. There might be different perspectives from other populations. Second, our studies were conducted in the US with participants who were familiar with ASL as sign language. The results may be different for learners with other backgrounds. Third, our DHH participants were all from a university specifically for the DHH community. Other DHH learners who are in colleges for all learners may have different perspectives toward emotional and online learning experiences. Fourth, our current prototype presented an emotion legend with text labels around nine emojis that participants may choose from to make self-reports. While our participants did not find these text labels distracting or misleading, future studies should investigate whether providing different or additional emojis to choose from, as well as whether or not providing keywords, may impact their self-reported emotion. 

Our work is grounded in the theory, \textit{Inclusive Special Education}, that mainstream education interventions can be adapted to include as many learners as possible, emphasizing SRL for post-school participation in society. However, other approaches exist, such as co-designing with special educators from scratch or using a mix of special and general education theories to guide technology design. Alternatively, one could start with general technology and add plug-ins to accommodate more students. While current HCI and CSCW works do not always ground accessible learning technology designs in this way, we encourage researchers to consider it. Additionally, we aimed to minimize the feeling of "being studied" among our DHH participants by encouraging them to understand their feelings and make suggestions. This approach helped them articulate their emotional journeys and discuss the educational challenges they have faced, which might have otherwise gone unrecognized.


\section{Conclusions}


Informed by Self-Regulated Learning (SRL) theory, using it to increase emotion awareness in online learning has shown promise for learners. However, it is unknown how learners with disabilities, such as DHH learners, may benefit from such technology. We conducted a comparative analysis between 20 DHH college students and 20 hearing college students to understand how and why DHH learners cannot benefit similarly from the prototype for reflecting on self-reported emotion and AER in video-based learning. We found that language diversity inhibited DHH students' ability to fully describe visualization and demonstrate their reflection process. Additionally, DHH participants were often too confused to describe their emotions in self-reporting without rewatching the videos. DHH participants also expressed concerns about the accuracy and privacy of using AER technology. They were interested in peers' emotions, and informed by this idea, we conducted an analysis of AER to find that DHH participants showed more positive emotions when viewing non-talking-head segments of the video. We discussed how SRL theory applies to learners with disability. For example, DHH participants expressed reflected sentences due to language diversity. Our work is grounded in \textit{Inclusive Special Education}, which aims to ensure effective education values mainstream education intervention to some extent and facilitate the highest level of inclusion in society post-school. This approach is one way, but not the only or best way, to approach learning technology design for learners with disabilities.



\bibliographystyle{ACM-Reference-Format}


\clearpage
\appendix
\renewcommand{\thefigure}{A.\arabic{figure}}
\setcounter{figure}{0}

\onecolumn
\section*{Appendix}
\label{appendix}



\section{Survey Questions} \label{appendix: survey}

The survey questions were adapted based on prior work. For each question, we used an 1-7 Likert scale with 1 as strongly disagree, 2 as disagree, 3 as moderately disagree, 4 as neutral, 5 as moderately agree, 6 as agree, and 7 as strongly agree.

\begin{enumerate}

    \item Navigating peers' emotions/cues increases my privacy concerns  than only seeing my own emotion
    \item Navigating peers' emotions/cues help me notice the limitations or the tool than only seeing my own emotion
    \item Navigating peers' emotions/cues help me notice the actual evidence behind the tool than only seeing my own emotion 
    \item Navigating peers' emotions/cues increases my trust in the tool than only seeing my own emotion
    \item Navigating peers' emotions/cues decreases my trust in the tool than only seeing my own emotion

\end{enumerate}

\clearpage

\end{document}